\documentclass[aps,showpacs,nofootinbib,twocolumn]{revtex4}

\usepackage{latexsym}
\usepackage{epsfig}
\usepackage{amsmath}
\usepackage{amssymb}
\usepackage{graphicx}

\begin{document}

%-----------------------------------------------------------------------

%\title{New asymptotically flat wormhole solutions with phantom equation of state }
\title{New asymptotically flat phantom wormhole solutions}

\author{Francisco S. N. Lobo$^{1}$}\email{flobo@cii.fc.ul.pt}
\author{Foad Parsaei$^{2}$}
\author{Nematollah Riazi$^{2.3}$}\email{riazi@physics.susc.ac.ir}

\affiliation{$^{1}$Centro de Astronomia e Astrof\'{\i}sica da
Universidade de Lisboa, Campo Grande, Edif\'{i}cio C8 1749-016
Lisboa, Portugal} \affiliation{$^{2}$Physics Department and Biruni
Observatory, Shiraz University, Shiraz 71454, Iran}
\affiliation{$^{3}$ Physics Department, Shahid Beheshti University,
Tehran 19839, Iran}

\date{\today}

%-----------------------------------------------------------------------

\begin{abstract}

A possible cause of the late-time cosmic acceleration is an exotic fluid with an equation of state lying within the phantom regime, i.e., $w=p/\rho <-1$. The latter violates the null energy condition, which is a fundamental ingredient in wormhole physics. Thus, cosmic phantom energy may, in principle, provide a natural fluid to support wormholes. In this work, we find new asymptotically flat wormhole solutions supported by the phantom energy equation of state, consequently extending previous solutions. Thus, there is no need to surgically paste the interior wormhole geometry to an exterior vacuum spacetime. In the first example, we carefully construct a specific shape function, where the energy density and pressures vanish at large distances as $\sim 1/r^{n}$, with $n>0$. We also consider the ``volume integral quantifier'', which provides useful information regarding the total amount of energy condition violating matter, and show that, in principle, it is possible to construct asymptotically flat wormhole solutions with an arbitrary small amount of energy condition violating matter. In the second example, we analyse two equations of state, i.e., $p_r=p_r(\rho)$ and $p_t=p_t(\rho)$, where we consider a specific integrability condition in order to obtain exact asymptotically flat wormhole solutions. In the final example, we postulate a smooth energy density profile, possessing a maximum at the throat and vanishing at spatial infinity.

%\textbf{Keywords} : wormholes, phantom energy, energy conditions

\end{abstract}

\maketitle

\section{Introduction}

One of the most intriguing discoveries in contemporary cosmology is that of
the accelerated expansion of the Universe~\cite{1}. A dark energy
component which dominates the Universe and has an equation of
state $p=w\rho$ with $w<-1/3$ is thought to be responsible for
this accelerated expansion. The specific case of $w<-1$ is denoted as the phantom energy
equation of state and although exotic in the laboratory context, is by
no means excluded by cosmological observations~\cite{2}.
Phantom energy possesses peculiar features, such as, a divergent cosmic energy density in a finite
time~\cite{cald}, it predicts the existence of a new long range
force \cite{15}, and leads to the appearance of a negative entropy and
negative temperature \cite{16}. Indeed, quintessence models which
exploit a real scalar field $\phi(t)$ with a self-interaction
potential $V(\phi)$ could properly play the role of a negative
pressure fluid, provided that the time variation rate of the
scalar field is much less than the field potential energy (i.e.
$\dot{\phi}^2 \ll V(\phi)$)\cite{a3}.

As the phantom equation of state parameter lies in the region
$w<-1$, it violates the null energy condition, i.e.,
$T_{\mu\nu}k^{\mu} k^{\nu} \geq 0 $, where $k^{\mu}$ is any null
vector. A stress-energy tensor $T_{\mu\nu}$ that violates the null
energy condition is denoted {\it exotic matter}. Thus, it seems that
phantom energy may in principle provide a means to support
traversable wormholes geometries~\cite{phantomWH,phantomWH2}.
Indeed, a fundamental ingredient of wormhole geometries is the
violation of the null energy condition \cite{WH},
which guarantees the flaring-out of the wormhole throat. In fact,
traversable wormholes violate all of the pointwise energy conditions
and the averaged energy conditions. In this context, the violation
of the pointwise energy conditions led to the averaging of the
energy conditions over timelike or null geodesics \cite{Tipler}. The
averaged energy conditions permit localized violations of the energy
conditions, as long on average the energy conditions hold when
integrated along timelike or null geodesics (see \cite{Lobo:2007zb}
for a recent review). The averaged energy conditions
involve a line integral, with dimensions (mass)/(area), and not a
volume integral, and therefore do not provide useful information
regarding the ``total amount'' of energy-condition violating matter.
Therefore, this prompted the proposal of  a ``volume integral
quantifier'' which amounts to calculating the following definite
integrals: $\int T_{\mu\nu}k^{\mu} k^{\nu}  dV$ and $\int
T_{\mu\nu}U^{\mu} U^{\nu}  dV$, where $U^\mu$ is the four-velocity
\cite{Visser}. The amount of energy condition violations is then the
extent that these integrals become negative.

Now, as the violation of the energy conditions is often viewed with
some scepticism, this led among other issues \cite{mty}, at least for some time, to the general
opinion that wormholes are physically implausible objects. However,
as mentioned above, the recent cosmological evidence for cosmic
energies which possibly violate the energy conditions has provided a
new motivation for taking wormhole solutions more seriously, at
least on theoretical grounds. Thus, the
phantom cosmic fluid presents a natural scenario for the
existence of these exotic geometries. Indeed, due to the fact of the
accelerating Universe, macroscopic wormholes could naturally be
grown from the submicroscopic constructions that originally pervaded
the quantum foam. One could also imagine an absurdly advanced civilization
mining the cosmic fluid for phantom energy necessary to construct
and sustain a traversable wormhole \cite{phantomWH2}.
However, many of the  papers published on wormholes supported by phantom energy are not asymptotically flat \cite{phantomWH,phantomWH2}. The interior wormhole metric is glued to a vacuum exterior spacetime at a junction interface \cite{thinshellwh}.

Indeed, as the phantom energy equation of state represents a spatially homogeneous cosmic fluid and is assumed not to cluster, it is also possible that inhomogeneities may arise due to gravitational instabilities. Thus, phantom energy wormholes may have originated from density fluctuations in the cosmological background, resulting in the nucleation through the respective density perturbations. Therefore, the pressure in the equation of state may be regarded as a radial negative pressure, and the tangential pressure is deduced through the Einstein field equations. Once these phantom wormholes arise from the density fluctuations, one can also consider the possibility that these structures be sustained by their own quantum fluctuations \cite{Garattini:2007ff}. 

Another interesting possibility analysed in the literature, is to consider a time-dependent dark energy star model \cite{Lobo:2005uf}, with an evolving parameter $w$ crossing the phantom divide  \cite{DeBenedictis:2008qm}. Once in the phantom regime, the null energy condition is violated, which physically implies that the negative radial pressure exceeds the energy density. Therefore, an enormous negative pressure in the center may, in principle, imply a topology change, consequently opening up a tunnel and converting the dark energy star into a wormhole  \cite{DeBenedictis:2008qm}. It is also possible to distinguish wormhole geometries by using astrophysical observations of the emission spectra from accretion disks, as specific signatures appear in the electromagnetic spectrum of accretion disks around wormhole geometries \cite{observationalWH}.

Recently, an alternative approach to wormhole physics has been explored within the context of modified gravity. It was shown that, in principle, it is possible to impose that the normal matter threading the wormhole satisfies the energy conditions, and it is the higher order curvature terms that support these exotic spacetimes \cite{Harko:2013yb}. Note that in modified gravity, the gravitational field equation may be expressed as an effective Einstein field equation, i.e., $G_{\mu\nu}\equiv R_{\mu\nu}-\frac{1}{2}R\,g_{\mu\nu}= \kappa^2 T^{{\rm eff}}_{\mu\nu}$, where $ T^{{\rm eff}}_{\mu\nu}$ is the effective stress-energy tensor, which contains the higher order curvature terms. Therefore, in modified theories of gravity, it is the effective stress-energy tensor that violates the null energy condition, i.e., $T^{{\rm eff}}_{\mu\nu} k^\mu k^\nu < 0$. Indeed, this approach has been extensively analysed in the literature, namely, in $f(R)$ gravity \cite{modgravity1}, curvature-matter couplings \cite{modgravity2}, conformal Weyl gravity \cite{modgravity3}, in braneworlds \cite{modgravity4}, and solutions \cite{Capozziello:2012hr} in the recently proposed hybrid metric-Palatini gravitational theory \cite{Harko:2011nh}, among other contexts.

In this work, we are primarily interested in presenting new asymptotically flat traversable wormholes governed by the phantom energy equation of state. The organization of the paper is as follows: In Section \ref{sec2}, the general equations and conditions in wormhole physics are briefly outlined. In Section \ref{sec3}, new solutions which are asymptotically flat are presented, the gravitational redshift caused by the specific parameters of the models are analysed, and the ``volume integral quantifier'' in several of the models is explored. In Section \ref{section4}, alternative approaches are outlined, in particular, we consider an equation of state relating the tangential pressure and the energy density, and in the second example we postulate a smooth energy density distribution. Finally, in Section \ref{sec:concl}, we present our concluding remarks. Throughout this work, we consider geometrised units, $G=c=1$.

\section{Basic formulation of wormhole theory}\label{sec2}

Wormholes are hypothetical geometrical structures connecting two
universes or two distant parts of the same universe \citep{WH}.  The general
static and spherically symmetric line element representing a wormhole is given by
\begin{equation}\label{1}
ds^2=-U(r)dt^2+\left[ 1-\frac{b(r)}{r} \right] ^{-1} dr^2+r^2\,d\Omega^2 \,,
\end{equation}
where $d\Omega^2= (d\theta^2+\sin^2\theta d\phi^2)$.  The metric function $b(r)$ is denoted the shape function, as it is related to the shape of the wormhole. The wormhole throat, which connects two asymptotic regions, is located at a minimum radial coordinate $r_0$, with $b(r_0)=r_0$. The shape function $b(r)$ must satisfy the so-called {\it flaring-out condition} given by $(b-b' r)/2b^2 > 0$, which reduces to $b'(r_0)<1$ at the wormhole throat. The condition $(1-b/r)>0$ is also imposed, so that $b(r)<r$, for $r>r_0$.

The metric function $U(r)> 0$ is denoted by the redshift function, as it is related to the gravitational redshift. More specifically, consider the gravitational redshift of a signal emitted at the wormhole throat and detected at an arbitrary large radial coordinate, so that
\begin{equation}
\frac{\tau(r)}{\tau(r_0)} =\frac{\lambda(r)}{\lambda
(r_0)}=\sqrt{\frac{g_{00}(r)}{g_{00}(r_0)}}\,,
\end{equation}
where $\tau(r)$ is the proper time of an observer located at a specific radial coordinate $r$. Thus, the redshift, as measured by a distant observer is given by
\begin{equation}
\frac{\lambda (r\rightarrow \infty
)}{\lambda(r=r_0)}=\frac{1}{\sqrt{U(r_0)}}\,,
\end{equation}
or equivalently by $z=\delta \lambda/\lambda= 1- 1/\sqrt{U(r_0)}$.

In this work, we are interested in obtaining asymptotically flat geometries. Thus, the metric functions need to obey the following conditions $U(r) \rightarrow 1 $ and $b(r)/r \rightarrow 0$, at $r\rightarrow \infty$.

We consider an anisotropic fluid in the form $T^\mu_\nu={\rm diag}(-\rho, p_r,p_t,p_t )$ to be the matter content of the spacetime, where $\rho(r)$ is the energy density, $p_r(r)$ is the radial pressure, and $p_t(r)$ is the lateral pressure measured in the orthogonal direction to the radial direction. The Einstein field equations provide the following distribution of matter 
\begin{eqnarray}\label{bprime2}
b'&=&8\pi r^2\rho, \\
\label{6}
\frac{U'}{U}&=&\frac{8\pi p_r r^3+b}{r(r-b)},\\
p_t&=&p_r+\frac{r}{2}\left[ p'_r + (\rho + p_r) \frac{U'}{2U}  \right]\,.
\label{pt}
\end{eqnarray}
Equation (\ref{pt}) may also be obtained using the conservation of the stress-energy tensor, $T^{\mu\nu}{}_{;\mu}=0$.

It is also very useful to rewrite the above Einstein field equations (\ref{bprime2})-(\ref{pt}) in terms of the stress-energy profile, given by by the following relationships
\begin{eqnarray}\label{bprime2b}
\rho(r)&=& \frac{1}{8\pi} \frac{b'}{r^2}, \\
\label{6b}
p_r(r)&=& \frac{1}{8\pi} \left[ \left(1- \frac{b}{r} \right)  \frac{U'}{U} - \frac{b}{r^3} \right],\\
p_t(r)&=&\frac{1}{16\pi} \left(1- \frac{b}{r} \right) \Bigg[   \frac{U''}{U} + \frac{U'}{rU} 
     \nonumber  \\
&&- \frac{b'r-b}{2r(r-b)} \frac{U'}{U} - \frac{b'r-b}{r^2(r-b)} \Bigg]\,.
\label{ptb}
\end{eqnarray}

Note that a mass function $m(r)$ can be defined according to
\begin{equation}\label{7}
m(r)\equiv \int_{r_0}^r 4\pi r^2\rho \,dr,
\end{equation}
which together with Eq. (\ref{bprime2}) leads to
\begin{equation}\label{8}
b(r)=r_0+2m(r).
\end{equation}

Thus, in this paper, we will be interested in constructing wormhole solutions using the equation of state $p_r(r)=w \rho(r)$, with $w<-1$. Taking into account the Einstein field equations (\ref{bprime2})-(\ref{6}), we have the following ordinary differential equation
\begin{equation}
\frac{U'}{U}=\frac{b+wrb'}{r^2(1-b/r)}\,.
  \label{ode}
\end{equation}

It is important to emphasize a subtlety in considering the phantom energy equation of state in the context of an inhomogeneous spherically symmetric spacetime. Note that phantom energy is a homogeneously distributed fluid, with an isotropic pressure, as emphasized in \cite{phantomWH,phantomWH2}. However, it can be extended to an inhomogeneous wormhole spacetime by assuming that the pressure in the equation of state is a negative radial pressure, and the transverse pressure may be determined from the Einstein field equation (\ref{pt}).
This is motivated by the discussion of the inhomogeneities that may arise due to gravitational instabilities, considered in the Introduction, and through the analysis carried out in \cite{SKim} where a 
time-dependent solution describing a spherically symmetric wormhole in a cosmological setting with a ghost scalar field was explored. More specifically, it was shown that the radial pressure is negative throughout the spacetime, and for large values of the radial coordinate, equals the lateral pressure, which demonstrates that the ghost scalar field behaves essentially as dark energy.

We now have several strategies to solve the field equations. We have four equations, Eqs. (\ref{bprime2})-(\ref{pt}) and (\ref{ode}), with five unknown functions, namely, $U(r)$, $b(r)$, $\rho(r)$, $p_r(r)$ and $p_t(r)$. To close the system, one may consider an extra equation of state $p_t=p_t(\rho)$, or alternatively, it is possible to consider specific choices for the distribution of the energy density threading the wormhole, much in the spirit of \cite{phantomWH, Dymnikova:2003vt}, or in specifying by hand the redshift function or the shape function, i.e., $U(r)$ and $b(r)$, respectively. Another possibility in obtaining a well-defined solution would be in finding an additional (differential) equation, that could be motivated by some appropriate variational principle, as discussed in 
\cite{Cattoen:2005he}. It was argued that whether the additional differential equation be an equation of state for the transverse pressure or a variational principle for the total anisotropy, it should be responsible for maintaining stability of the solution over a wide range of total masses and central pressures. Consequently, the closed set of equations must be self-regulating if it is to be physically interesting \cite{Cattoen:2005he}. 

In this work, we are primarily motivated in finding asymptotically flat wormhole solutions, and consequently consider a variety of the above approaches. More specifically, first we choose a specific shape function and find asymptotically flat wormhole solutions that are supported by an arbitrary small amount of energy condition violating matter, as provided by the ``volume integral quantifier''. In second place, we obtain further asymptotically flat wormhole geometries, by considering two specific approaches, namely, (i) a second equation of state relating the tangential pressure and the energy density, and (ii) by considering a smooth energy density profile, containing a maximum at the throat, and dropping off to zero at spatial infinity.

\section{Asymptotically flat phantom wormhole solutions I: Specifying the shape function}\label{sec3}

Throughout this section, we consider wormholes with the following shape function
\begin{equation}\label{10}
\frac{b(r)}{r_0}=a\left( \frac{r}{r_0}\right)^\alpha +C,
\end{equation}
where $a$, $\alpha$, and $C$ are dimensionless constants.

We will show below that the parameter $a$ will play a fundamental role in determining the gravitational redshift of signals, and in obtaining arbitrary small quantities of energy condition violating matter by evaluating the ``volume integral quantifier'' \cite{Visser}.

Note that in order to obey the condition $b(r)/r \rightarrow 0$, we need to impose that $\alpha<1$. Evaluated at the throat, i.e., $b(r_0)/r_0=a+C=1$, one finds that the constant is given by $C=1-a$. Thus, the shape function finally takes the form
\begin{equation}
b(r)=r_0+ar_0\left[ \left( \frac{r}{r_0} \right)^{\alpha} -1\right] \,.
  \label{shape}
\end{equation}
Considering a positive energy density yields the imposition $a \alpha >0$. In order to satisfy the flaring-out condition at the throat, $b'(r_0)<1$, one deduces the additional constraint $ a\alpha <1$. In summary, one has the following restrictions for the parameters
\begin{equation}
\alpha < 1, \qquad 0< a \alpha <1 \,.
\end{equation}

The condition of an event-horizon-free spacetime requires that $U(r)$ be finite and non-zero. Thus, due to the finiteness of $U(r)$, the radial pressure evaluated at throat is given by $p_r(r_0)=-1/(8\pi r_0^2)$ (see Eq. (\ref{6})). The latter negative radial pressure at the throat provides the geometry with a repulsive character, preventing the wormhole from collapsing \cite{WH}. Now, the equation of state $p_r=w \rho$, at the throat, also imposes the following condition
\begin{equation}\label{15}
aw\alpha=-1.
\end{equation}

In addition to the conditions imposed above, one also needs to consider that $(1-b/r)>0$, or equivalently $b(r)<r$. This condition may be expressed by the following inequality
\begin{equation}\label{16}
H(x,a) \equiv a \left(\frac{r}{r_0}\right)^\alpha-\frac{r}{r_0} +1-a<0,
\end{equation}
where $x\equiv r_0/r$. The latter parameter range has been defined, in order to define the entire spacetime, and has the range $0<x \leq 1$. Note that $x=1$ corresponds to the wormhole throat, $r=r_0$, and $x \rightarrow 0$ to spatial infinity $r\rightarrow \infty$. In order to check this condition for various choices of $\alpha$ and $a$, we shall plot $H(x,a)$ against $x= r_0/r$ and $a$ for specific choices of the parameter $\alpha$, for the solutions obtained below.

The strategy that is adopted in this section is the following. Considering the shape function given by Eq. (\ref{shape}), the ordinary differential equation, given by Eq. (\ref{ode}), takes the following form
\begin{equation}
\frac{U'}{U}=\left(\frac{r_0}{r^2}\right)\frac{1+a \left[  \left(\frac{r}{r_0} \right)^\alpha (1+w\alpha) -1\right] }{1-\left( \frac{r_0}{r}\right)\left\{ 1+a \left[ \left( \frac{r}{r_0}\right)^\alpha -1 \right]  \right\} } \,, \label{ode2}
\end{equation}
which does not -- in general -- yield an exact solution. Thus, in this
section, in order to deduce exact wormhole solutions of Eq.
(\ref{ode2}), we will impose specific choices for the parameters $a$
and $\alpha$, which will be considered below.

\subsection{Wormholes with vanishing redshift function}\label{subsec1}

The first case of interest is that of $a=1$, so that $\alpha w=-1$, and taking into account Eq. (\ref{ode2}), leads to the solution $U=U_0={\rm constant}$. In this case, all of the wormhole conditions discussed in the previous section are satisfied, provided that $\alpha$ is in the range $0<\alpha<1$.

For this case, the line element becomes
\begin{equation}\label{18}
ds^2=-dt^2+\frac{dr^2}{1-(r_0/r)^{1-\alpha}}+r^2d\Omega^2\,,
\end{equation}
where we have absorbed the factor $U_0$ into the re-scaled time coordinate $t$.

The stress-energy tensor profile is given by
\begin{eqnarray}\label{19}
p_r(r)&=&w\rho(r)=-\frac{1}{8\pi r_0^2}\left(\frac{r_0}{r}\right)^{3-\alpha}\,,\\
p_t(r)&=& \frac{\alpha -1}{2}p_r(r).
\end{eqnarray}
This case which represents a simple phantom wormhole with a vanishing redshift function was previously considered in \cite{rah}, so that we will not discuss it further here.

\subsection{Wormholes with an unbounded mass function}\label{subsec2}

\subsubsection{Specific case: $\alpha=1/2$}

For the specific case  $\alpha=1/2$, the shape function takes the form
\begin{equation}\label{22}
\frac{b(r)}{r_0}=a\sqrt{\frac{r}{r_0}}-a+1 \,,
\end{equation}
so that the mass function reads
\begin{equation}\label{27}
m(r)= \frac{a}{2}\left( \sqrt{rr_0}-r_0\right)\,,
\end{equation}
which is positive throughout the spacetime, but unbounded as $r\rightarrow \infty$.

Taking into account the shape function given by Eq. (\ref{22}), the ordinary differential equation (\ref{ode2}) yields the solution
\begin{equation}\label{21}
U(r)=k\left(1+\frac{1-a}{\sqrt{r/r_0}}\right)^{2},
\end{equation}
where $k$ is  a constant of integration.

The line element then takes the form
\begin{equation}\label{23}
ds^2=-\left(1+\frac{1-a}{\sqrt{r/r_0}}\right)^{2}dt^2+\frac{dr^2}{1-\frac{a}{\sqrt{r/r_0}}-\frac{1-a}{r/r_0}}+r^2\,d\Omega^2\,,
\end{equation}
where the factor $k$ has been absorbed into a redefinition of the time coordinate.

Note that for this case, we have $w=-2/a$, and $a$ lies in the range $0<a<2$. It is seen that the wormhole spacetime is asymptotically flat, and there is no event horizon, since $U$ is finite and non-zero throughout the spacetime.

\begin{figure}[h]
\includegraphics[width=3.8in]{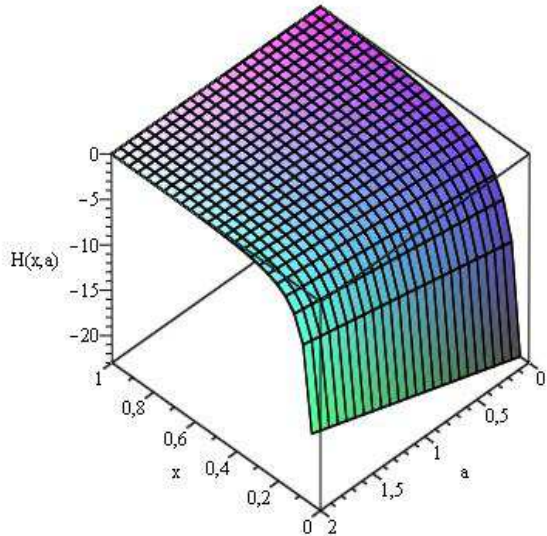}
\caption{The plot depicts the function $H(x,a)$, for $\alpha=1/2$ and where the parameter $x=r_0/r$, lying in the range $0<x \leq 1$, has been defined in order to define the entire spacetime. It is transparent that $H(x,a)$ is negative throughout the entire range of $x$. See the text for details.}\label{fig1}
\end{figure}

In order to check the condition $(1-b/r)>0$, or equivalently the inequality (\ref{16}), we have plotted $H(x,a)$ as a function of $x=r_0/r$ and $a$ in Fig. \ref{fig1}, where it is transparent that $H(x,a)$ is negative throughout the entire range of $x$.

The gravitational redshift of a signal emitted at the wormhole throat and detected by an observer at a large radial coordinate is given by
\begin{equation}\label{42}
z=\frac{\delta \lambda}{\lambda}=\frac{1-a}{2-a}\,.
\end{equation}
The parameter $a$, therefore, directly affects the redshift of signals originating from the throat or its neighbourhood. Note that the gravitational redshift is zero, for $a=1$. If $a \rightarrow 2$, then $z \rightarrow \infty$ which unveils an event horizon at $r \rightarrow r_0$.

The stress-energy tensor profile needed to support this spacetime is give by the following expressions:
\begin{eqnarray}\label{24}
p_r(r)&=&w\rho(r)=-\frac{1}{8\pi r_0^2} \left( \frac{r_0}{r} \right)^{3-\alpha},\\
p_t(r)&=&-\frac{1}{4}p_r(r)
   \nonumber  \\
&&\hspace{-1.5cm}+\frac{(r_0/r)^{1/2}(2-5a-a^3+4a^2)+(2-3a+a^2)}{64\pi
r^{2}[(r/r_0)^{1/2}+1-a]^2}.
\end{eqnarray}
Note that although the phantom fluid is anisotropic, it becomes asymptotically isotropic at large
$r$.

\subsubsection{Specific case: $a=\frac{1}{3}$ and
$\alpha=\frac{1}{3}$}

Another case for which  we may find $U(r)$ analytically is considering $a=\frac{1}{3}$ and
$\alpha=\frac{1}{3}$, so that Eq. (\ref{ode2}) yields the following solution
\begin{eqnarray}\label{29}
U(r)&=&k\left[3+3\left(\frac{r_0}{r}\right)^{\frac{1}{3}}+2\left(\frac{r_0}{r}\right)^{\frac{2}{3}}\right]^{\frac{3}{2}}
   \nonumber  \\
&&\hspace{-1cm}\times\exp\left\{
\frac{3\sqrt{15}}{5}\arctan\left[\frac{\sqrt{15}(2(r/r_0)^{\frac{1}{3}}+1)}{5}\right]\right\}.
\end{eqnarray}
Thus, the line element becomes
\begin{equation}\label{30}
ds^2=-U(r)dt^2+\frac{dr^2}{1-\frac{1}{3} \left(  \frac{r_0}{r} \right)^{\frac{2}{3}} -\frac{2r_0}{3r}} +r^2d\Omega^2 \,.
\end{equation}

The stress-energy tensor components that thread the wormhole, are given by
\begin{eqnarray}\label{31}
p_r(r)&=&w\rho(r)=-\frac{1}{8\pi r_0^2}   \left(  \frac{r_0}{r}
\right)^{\frac{8}{3}} ,
\end{eqnarray}
and
\begin{equation}
p_t= - \frac{1}{3}p_r +\frac{1}{18\pi r^2
(r/r_0)^{2/3}\left[2+3(r/r_0)^{2/3}+3(r/r_0)^{1/3}\right]}.
\end{equation}

In order to impose an asymptotically flat spacetime, i.e., with $U(r)\rightarrow 1$ for $r\rightarrow +\infty$, the constant of integration takes the value $k= 3^{-2/3}\left(e^{\pi \sqrt{15} }\right)^{-3/10}$.

It is also interesting to evaluate the ``volume integral quantifier'' \cite{Visser}, referred to in the Introduction, and which is given by
\begin{eqnarray}
I_V\equiv \int (\rho+p_r)dV&=&\left[ (r-b)\ln \left(\frac{U(r)}{1-b/r}\right)\right]^\infty_{r_0}
   \nonumber  \\
&& \hspace{-1.5cm} - \int_{r_0}^\infty
[1-b'(r)]\left[ \ln \left(
\frac{U(r)}{1-\frac{b}{r}}\right)\right] dr.
\end{eqnarray}
The latter provides information about the ``total amount''  of
energy condition violating matter in the spacetime. For the present
solution, we verify that an infinite ``amount of energy conditions
violating matter'' is needed in order to have an asymptotically flat
wormhole geometry. More specifically,  the volume integral
quantifier tends to be globally violated, since $I_V=
(-2+a)r^{1/2}|^{\infty}_{r_0}\rightarrow -\infty$. Thus, this
motivates the search for a solution where the ``volume integral
quantifier'' is finite, which shall be analysed below.

\subsection{Wormholes with a bounded mass function}\label{subsec3}

An interesting solution corresponds to $\alpha=-1$. For this value, we verify that $aw=1$, so that for the phantom energy equation of state parameter, we have $-1< a < 0$.
It can be seen from Eq. (\ref{shape}) that for this value of
$\alpha$, there is a mass function with finite value in which the
parameter $a$ plays an important role
\begin{equation}\label{37}
m(r)= \frac{ar_0}{2}\left(\frac{r_0}{r} -1\right).
\end{equation}

Equation (\ref{ode2}) yields the following solution
\begin{equation}
U(r)=k\left(1+ \frac{ar_0}{r}  \right)^{1-1/a}\,,
\end{equation}
where $k$ is an a constant of integration.

\begin{figure}[h]
\includegraphics[width=3.8in]{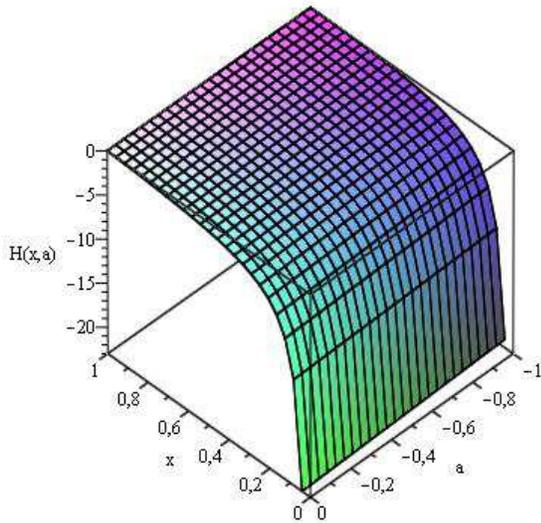}
\caption{The surface in the plot depicts the function $H(x,a)$ for $\alpha=-1$. The parameter $x=r_0/r$ has been defined and lies in the range $0<x \leq 1$, in order to define the entire spacetime. It is transparent that $H(x,a)$ is negative throughout the spacetime. See the text for details.}\label{fig2}
\end{figure}

In order to check  condition (\ref{16}), the function $H(x,a)$ is depicted as the surface in
Fig. \ref{fig2}, which is negative throughout the spacetime, so that the
condition $1-b(r)/r > 0$ is satisfied.

The line element in this case is given by
\begin{equation}
ds^2=-\left(1+\frac{ar_0}{r}\right)^{1-\frac{1}{a}}dt^2
+\frac{dr^2}{1-\frac{r_0}{r}\left(\frac{ar_0}{r}+1-a\right)}
+r^2\,d\Omega^2.
\end{equation}
where it is transparent that the geometry is asymptotically flat.

The gravitational redshift, as measured by a distant observer is given by
\begin{equation}
\frac{\lambda (r\rightarrow \infty
)}{\lambda(r=r_0)}=(1+a)^{\frac{1-a}{2a}},
\end{equation}
or
\begin{equation}\label{40}
z=\frac{\delta \lambda}{\lambda}=1-(1+a)^{\frac{1-a}{2a}}\,,
\end{equation}
Thus, the parameter $a$ directly affects the redshift of signals originating from the throat or its nearby
regions.

The energy density and pressure profile which supports this spacetime is given by the following expressions
\begin{eqnarray}\label{36}
p_r(r)&=&w\rho(r)=-\frac{1}{8\pi r_0^2}\left(\frac{r_0}{r} \right)^{4},\\
p_t(r)&=&
-\frac{1}{4}p_r+\frac{\left[r_0(a^2+3a-1)+3r\right]r_0^2}{32\pi
r^4(r+ar_0)}.
\end{eqnarray}

The ``volume integral quantifier'' \cite{Visser}, for this case is given by
\begin{equation}
I_V \equiv \int(\rho+p_r)dV =2\int_{r_0}^\infty (\rho +p_r)4\pi r^2 dr
=-(1+a)r_0\,.
\end{equation}
Note that for $a\rightarrow -1$, we have $I_V \rightarrow 0$, which reflects arbitrary small quantities of energy condition violating matter. Thus, by carefully constructing the shape function the specific shape function given by  Eq. (\ref{shape}), one may in principle construct asymptotically flat spacetimes supported by arbitrary small quantities of exotic matter.

\section{Other asymptotically flat phantom wormhole solutions II}\label{section4}

One may argue that the approach outlined and explored above, namely, in specifying a shape function lacks physical justification and motivation for the stress-energy tensor distribution. Thus, in this section we consider two alternative approaches that provide asymptotically flat wormhole solutions. The first approach consists in specifying a second equation of state relating the tangential pressure and the energy density. In the second example, we consider a physically motivated smooth energy density profile, consisting of a maximum at the throat and decreasing to zero at spatial infinity.

\subsection{Specifying two equations of state}\label{section4a}

The system of equations (\ref{bprime2b})-(\ref{ptb}) are closed by taking into account two equations of state, namely, $p_r=p_r(\rho)$ and $p_t=p_t(\rho)$. The radial phantom energy equation of state is given by $p_r(r)=w \rho(r)$, which using the field equations provides the ordinary differential equation given by Eq. (\ref{ode}). 

Now, as we are considering anisotropic pressures, then it is useful to consider the following dimensionless anisotropy parameter
\begin{equation}
\Delta= \frac{p_t-p_r}{\rho}\,.
  \label{anisotropy}
\end{equation}
considered extensively in \cite{Cattoen:2005he}. Assuming that $\rho>0$, it is interesting to note that $\rho \Delta/r$ represents a force due to the anisotropic nature of the phantom wormhole model. If $\Delta > 0$, i.e., $p_t>p_r$, the geometry is repulsive, being outward directed, and attractive if $\Delta < 0$, i.e., $p_t<p_r$.

Note that for isotropic pressures, $p_t=p_r$, i.e., $\Delta=0$, then the system of equations is closed, although one cannot find exact solutions for the resulting differential equation. Thus, as mentioned in \cite{Cattoen:2005he}, in the context of gravastars, in considering anisotropic pressures, it may be efficient in choosing an equation of state depending on $\Delta$. Indeed, it was shown that in order to generically avoid the formation of an event horizon, then the material threading the gravastar should be sensitive to the ``compactness'' factor $b(r)/r$, so that the anisotropy equation of state should have the following variable-dependence: $\Delta=\Delta(r,p_r,b/r)$. In the present context of phantom wormholes, we consider the particular case of $\Delta=\Delta(r)$, and taking into account Eq. (\ref{anisotropy}) implies an equation of state of the form
\begin{equation}
p_t=p_t(r,\rho)=\beta(r) \rho = \left[ \Delta(r) + w \right] \, \rho \,, 
  \label{tangEOS}
\end{equation}
where the parameter $\beta(r)$ depends on the dimensionless anisotropy factor and the phantom equation of state parameter, through the definition $\beta(r)=\Delta(r) + w$. 
Thus, the parameter $\beta(r)$ is denoted the dimensionless phantom anisotropy factor, and it will be useful in obtaining exact asymptotically flat phantom wormhole solutions, which is the prime concern of this paper. A similar approach was considered in \cite{Hille}. 

The equation of state (\ref{tangEOS}) relating the tangential pressure with energy density, yields the following differential equation
\begin{eqnarray}
\frac{2b' \beta(r)}{r^2}+\frac{(b'r-b)(b+wrb')}{2r^3(r-b)} + \frac{b'r-b}{r^3}
    \nonumber  \\
 =\left(  1-\frac{b}{r} \right) \left[ \frac{U''}{U} + \frac{1}{r} \frac{U'}{U}  \right] 
  \label{odeUr} \,,
\end{eqnarray}

Taking into account the ordinary differential equation, Eq. (\ref{ode}), provided by the radial phantom energy equation of state and Eq. (\ref{odeUr}), one arrives at the following nonlinear second order differential equation
\begin{eqnarray}
\left(  1-\frac{b}{r} \right) ^{-1}
\left[  \frac{2b' \beta(r)}{r^2}+\frac{(b'r-b)(b+wrb')}{2r^3(r-b)} + \frac{b'r-b}{r^3} \right]
    \nonumber  \\
 =\left[ \frac{b+wrb'}{r(r-b)} \right]'
 +
 \left( \frac{b+wrb'}{r^2} \right) \left[  \frac{b+wrb'}{r(r-b)}  +\frac{1}{r}  \right] \,. \label{nonlinearode}
\end{eqnarray}
One cannot find exact analytical solutions for this equation. Even the simplest case of a constant parameter does not yield an exact solution, although numerical solutions may be found. In this context, in order to obtain asymptotically flat phantom wormhole solutions, we consider a novel approach in that the $r$-dependent phantom anisotropy parameter $\beta(r)$ provides an extra degree of freedom. This latter feature imposes an integrability condition, which enables us to solve exactly  the differential equation (\ref{nonlinearode}).  A subtle issue is that in introducing an extra unknown function, namely, $\beta(r)$, the system becomes once again under-determined, however, the imposition of an integrability condition considered below closes once again the system of equations. 

In this context, we impose a particular integrability condition given by
\begin{eqnarray}
2b' \beta(r)+\frac{(b'r-b)(b+wrb')}{2r(r-b)} + \frac{b'r-b}{r}
    \nonumber  \\
 =\left(  1-\frac{b}{r} \right) \left[ \frac{\gamma \alpha^2 \left(\frac{r_0}{r} \right) ^\alpha }{1+\gamma \left( \frac{r_0}{r} \right)^\alpha } \right] \,,
\end{eqnarray}
so that Eq. (\ref{odeUr}) provides the following exact solution 
\begin{equation}
U(r)=\left[1+ \gamma \left( \frac{r_0}{r} \right)^\alpha \right] \,,
  \label{solutionU1}
\end{equation}
with $\alpha > 0$, so that $U \rightarrow 1$ at spatial infinity $r \rightarrow \infty$.

\begin{figure}[h]
\includegraphics[width=3.6in]{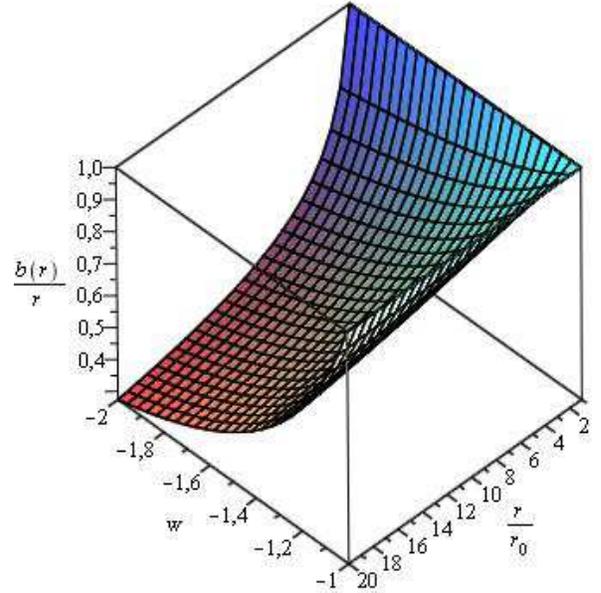}
\caption{The condition for $b(r)/r <1$ is satisfied throughout the spacetime, as is depicted in the plot. See the text for details.}\label{fig3}
\end{figure}

From Eq. (\ref{ode}), and considering $\alpha=1$, one finally arrives at the following shape function
\begin{equation}
b(r)=- \gamma r_0 + (r + \gamma r_0)^{-1/w}\,[r_0(1+\gamma)]^{(1+w)/w}\,.
\end{equation}
The radial derivative of the shape function is given by $b'(r)=-(1/w)(r + \gamma r_0)^{-(1+w)/w}\,[r_0(1+\gamma)]^{(1+w)/w}$, which evaluated at the throat reduces to $b'(r_0)=-1/w < 1$, consequently obeying the flaring out condition at the throat. The condition for $b(r)/r <1$ is also satisfied throughout the spacetime and is depicted in Fig. \ref{fig3}, for the specific value of $\gamma=-1/2$.

One may also analyse the attractive/repulsive nature of the geometry through the redshift function. To this effect, consider a static observer, at rest at constant $r,\theta, \phi$, with the four-velocity given by $u^{\mu}=dx^{\mu}/{d\tau}=(u^{\,t},0,0,0)=(1/\sqrt{U(r)},0,0,0)$. The observer's
four-acceleration is $a^{\mu} =u^{\mu}{}_{;\nu}\,u^{\nu}$, so that
taking into account the wormhole spacetime metric (\ref{1}), we arrive at $a^t=0$ and $a^r=(1-b/r)g$, where $g=U'/2U$ is defined for notational simplicity. Now, from the geodesic equation, a radially moving test particle, which initially starts at rest, obeys the following
equation of motion
\begin{equation}
\frac{d^2r}{d\tau^2}=-\Gamma^{r}{}_{tt}\left(\frac{dt}{d\tau}\right)^2=-a^r
\,.
      \label{radial-accel}
\end{equation}
The parameter $a^r$ is the radial component of proper acceleration that an observer must maintain in order to remain at rest at constant $r,\,\theta,\,\phi$. The geometry is attractive if $a^r>0$, i.e., an observer must maintain an outward-directed radial acceleration to keep from being pulled into the wormhole; and repulsive if $a^r<0$, i.e., an observer needs an inward-directed radial acceleration to avoid being pushed away from the wormhole. Note that this distinction depends on the sign of $g=U'/2U$. Thus, the convention used is that $g$ is positive for an inwardly gravitational attraction, and negative for an outward gravitational repulsion \cite{Roman:1992xj}.

For the specific solution obtained in Eq. (\ref{solutionU1}), we have 
\begin{equation}
g=\frac{U'(r)}{2U(r)}=-\frac{\gamma \alpha \left(\frac{r_0}{r}\right)^\alpha }{2r \left[ 1+\gamma \left( \frac{r_0}{r}  \right)^\alpha  \right] } 
\end{equation}
and taking into account the specific values of $\alpha=1$ and $\gamma=-1/2$, the factor reduces to $g=r_0/[4r^2(1-r_0/(2r))]$, which is positive throughout the spacetime, indicating an inwardly gravitational attractive nature of the geometry.

The stress-energy tensor profile is given by the following relationships
\begin{eqnarray}
p_r&=&w\rho =-\frac{1}{8\pi r^2}(r+\gamma r_0)^{-\frac{1+w}{w}}\left[r_0(1+\gamma)\right]^{\frac{1+w}{w}}\,,\\
p_t&=&\frac{1}{16\pi r^2} \Bigg\{ \left[  1- \frac{(r+\gamma r_0)^{-\frac{1}{w}}\left[r_0(1+\gamma)\right]^{\frac{1+w}{w}}-\gamma r_0}{r} \right] 
\nonumber  \\ 
 &&
  \Bigg[   \frac{\gamma r_0}{r(1+\frac{\gamma r_0}{r})}  
  + \frac{r_0(1+\gamma)(2r+\gamma r_0)}{2w(r+\gamma r_0)^2} \times
    \nonumber  \\
&& \hspace{-1.1cm}  \times \frac{\gamma w r(r+\gamma r_0)^{\frac{1}{w}}+\left[r_0(1+\gamma)\right]^{\frac{1+w}{w}}[r+w(r+\gamma r_0)]}{ r_0 \left[r_0(1+\gamma)\right]^{\frac{1+w}{w}}(1+\gamma)- \left(r+\gamma r_0\right)^{\frac{1+w}{w}}  }  \Bigg]  \Bigg\}\,.
\end{eqnarray}
The stress-energy components tend to zero at spatial infinity; and at the throat, $p_r|_{r_0}=w\rho|_{r_0}=-1/(8\pi r_0^2)$ and $p_t|_{r_0}=3(1+w)(2+\gamma)/[32\pi w r_0^2(1+\gamma)]$.

\begin{figure}[h]
\includegraphics[width=3.8in]{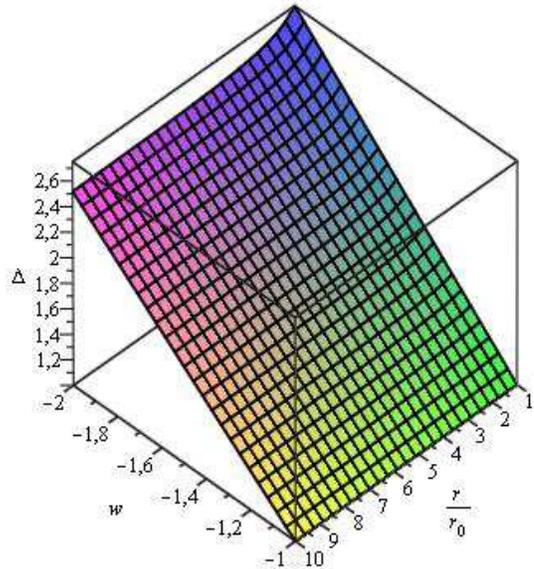}
\caption{The dimensionless anisotropy parameter is depicted for the values $\alpha=1$ and $\gamma=-1/2$. It is shown that $\Delta > 0$, which implies a repulsive character of the geometry due to the anisotropy of the system. See the text for details.}\label{fig:anisotropy}
\end{figure}

For this specific case, the dimensionless anisotropy parameter, given by Eq. (\ref{anisotropy}), is depicted in Fig. \ref{fig:anisotropy} for the values $\alpha=1$ and $\gamma=-1/2$. It is shown that $\Delta > 0$, which implies a repulsive character of the geometry due to the anisotropy of the system. Thus, one may naively conclude that the latter repulsive character due to the anisotropy compensates the attractive nature of the gravity profile for a specific range of the parameters of the wormhole model. This partially justifies the choice for the second equation of state that was considered. However, we emphasize that a full stability analysis of the system is in order.

\subsection{Smooth phantom energy density distribution}

One may also consider a specific energy density (or pressure) profile and close the system with a specific equation of state as done in \cite{phantomWH,Dymnikova:2003vt}. In this context, consider the specific case of the energy density profile given by
\begin{equation}
\rho(r)=\rho_0\left( \frac{r_0}{r} \right)^\alpha\,,
\end{equation}
where $\alpha >0$. Note that the energy density possesses a smooth profile with a maximum at the throat, i.e., $\rho_0$, and drops off to zero at spatial infinity. 

\begin{figure}[ht]
\includegraphics[width=3.8in]{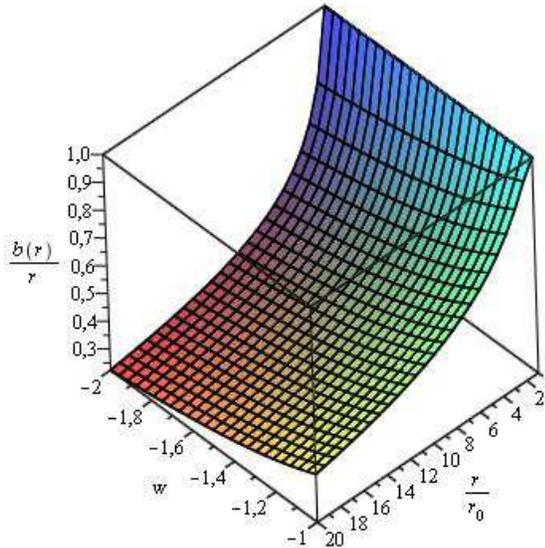}
\caption{The condition for $b(r)/r <1$ is satisfied throughout the spacetime, as is depicted in the plot. We have considered the following values for the parameters $\alpha=5/2$ and $\rho_0=-1/(8w \pi r_0^2)$. See the text for details.}\label{fig5a}
\end{figure}

From Eq. (\ref{bprime2}), one deduces the following shape function
\begin{equation}
b(r)=r_0-\frac{8\pi \rho_0 r_0^3}{3-\alpha} \left[ 1- \left( \frac{r_0}{r} \right) ^{\alpha-3} \right] \,.
\end{equation}
The radial derivative is given by $b'(r)=8\pi \rho_0 r^2(r_0/r)\alpha $, and at the throat reduces to $b'(r_0)=8\pi \rho_0 r_0^2 < 1$, which implies the following restriction $\rho_0<1/(8\pi \rho_0)$. In order  to obey the condition $b(r)/r \rightarrow 0$ at $r\rightarrow \infty$, the condition $\alpha >2 $ is also imposed.
Indeed, the condition for $b(r)/r <1$ is also satisfied throughout the spacetime and is depicted in Fig. \ref{fig5a}, where we have considered the following values for the parameters $\alpha=5/2$ and $\rho_0=-1/(8w \pi r_0^2)$.

\begin{figure}[h]
\includegraphics[width=2.6in]{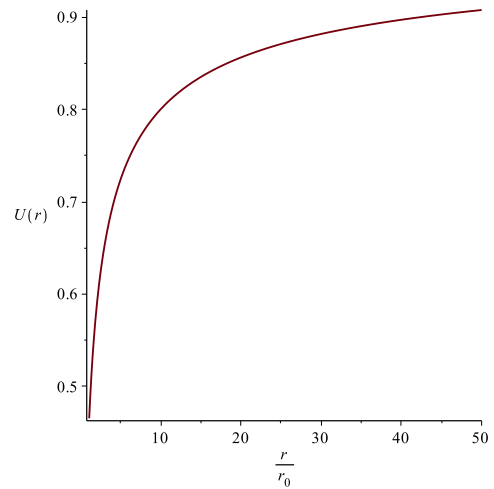}
\caption{Profile of the redshift function given by Eq. (\ref{redshiftfunction2}), for the spefific values of $\alpha=5/2$, $w=-1.5$ and $\rho_0=-1/(8\pi w r_0^2$. See the text for details.}\label{fig5b}
\end{figure}

One cannot find an exact general solution for the redshift function $U(r)$, with an arbitrary $\alpha$, but consider for simplicity the specific case $\alpha=5/2$, so that Eq. (\ref{ode}) yields the following solution
\begin{eqnarray}
U(r)&=& C \left[ \bar{\rho}_0  \left( \frac{r_0}{r} \right)\left(r^2 - r_0^2\right)+ \frac{r_0}{r}  -1 \right]  \times
    \nonumber \\
&& \hspace{-1.65cm} \exp\left\{   \frac{\bar{\rho}_0 r_0^{2}(1+w)\, {\rm arctanh}\left[  \frac{\sqrt{\frac{r}{r_0}} \left[ -1+\frac{\bar{\rho}_0}{2} r^2 
 \left( \frac{r_0}{r} \right)^{5/2} \right] }{ \sqrt{1 -\bar{\rho}_0 r_0^2 + \frac{\bar{\rho}_0^2 r_0^4}{4} } }  
 \right]  }{\sqrt{1 -\bar{\rho}_0 r_0^2 + \frac{\bar{\rho}_0^2 r_0^4}{4} }}   \right\}   \,,
   \label{redshiftfunction2}
\end{eqnarray}
where $\bar{\rho}_0=16\pi \rho_0$ is defined for notational convenience. 
The integration constant $C$ can be absorbed into a redefinition of the time coordinate. The latter function is depicted in Fig. \ref{fig5b}, where we have considered the following values for the parameters $w=-1.5$ and $\rho_0=-1/(8w \pi r_0^2)$.

Relative to the stress-energy tensor components, the tangential pressure is not presented as it is extremely messy and lengthy. However, it is possible to show that it is finite throughout the spacetime and does indeed tend to zero at spatial infinity.

\section{Concluding remarks}\label{sec:concl}

The possibility that the accelerated expansion of the Universe may be driven by an exotic fluid with an equation of state given by $w=p/\rho <-1$, denoted by phantom energy, is not ruled out by cosmological observations. Such a fluid violates the null energy condition, which is a fundamental ingredient to sustain static traversable wormholes. The possibility of phantom energy has therefore brought new life to research in wormhole physics, while opening fascinating observational possibilities through future observations of CMB and distant supernovae. In the present work, we have found new exact wormhole solutions supported by phantom energy. By carefully constructing a specific shape function, we have obtained the metric, as well as the energy density and pressure profile needed to support the wormhole geometries and discussed some of the properties of the resulting spacetime. In particular, the mass function of the wormhole was shown to be finite or infinite depending on the particular choices of the parameters. Despite this,  it was also shown that a signal emitted at the throat undergoes a finite redshift as measured by a distant observer, depending on the specific parameters of the theory. Furthermore, we have also explored alternative approaches, in particular, by (i) imposing an equation of state relating the tangential pressure and the energy density and (ii) imposing a smooth energy density distribution.

As compared with many of the solutions previously obtained in the literature, the line elements obtained in this work are asymptotically flat with an energy density and pressures vanishing at large distances from the wormhole as $\sim 1/r^{n}$, with $n>0$. The possibility of choosing different values for the parameters of the model, helped us find  exact solutions with special mass and redshift
values as seen by a distant observer. One important property of the solutions presented in this paper is that  there is no need to use surgeries to glue the wormhole to an exterior vacuum geometry. We have also considered the ``volume integral quantifier'', which provides useful information regarding the total amount of energy condition violating matter, and we have shown that, in principle, it is possible to construct asymptotically flat wormhole solutions with an arbitrary small amount of energy condition violating matter. 
As in the previously obtained solutions, these phantom energy traversable wormholes have far-reaching physical implications, such as, absurdly advanced civilizations may use these geometries for interstellar travel and to induce closed timelike curves, consequently violating causality.

Nevertheless, as there is no observational evidence pointing to the existence of wormholes, and as phantom energy represents a spatially homogeneous cosmic fluid and is assumed not to cluster, a qualitative argument regarding the stability of the system is in order. Indeed, as mentioned in the Introduction it is possible that inhomogeneities may arise due to gravitational instabilities, and that phantom energy wormholes may have originated from density fluctuations in the cosmological background, resulting in the nucleation through the respective density perturbations. In this context, a full-fledged stability analysis is in order. 
In fact, wormholes have been modelled by so-called ghost or phantom scalar fields, involving a simple Lagrangian describing a minimally coupled massless scalar field with a reversed sign kinetic term. The stability issue has been extensively explored in the literature, and in particular, it has been shown that massless non-minimally coupled scalar fields are unstable under linear \cite{Bronnikov} and nonlinear perturbations \cite{Gonzalez:2008xk}.
Recently, in order to obtain stable solutions, the study of configurations supported by a ghost scalar field with a self-interaction potential was analysed \cite{Dzhunushaliev:2013lna}. In particular, a ghost scalar field with a quartic coupling was analysed, and it was found that all the wormhole systems considered were unstable against linear perturbations. In fact, two serious difficulties seem to stand out, namely, instabilities arise at boundary surfaces separating ghost and normal fields, which generally transform these surfaces into singular ones \cite{instability1}, and in second place, ghost fields are extremely problematic at the quantum level. In the latter feature, the negative kinetic term leads to the possibility that the energy density may become arbitrarily negative for high frequency oscillations \cite{Sushkov:2007me}. It may be possible that a more general self-interaction potential, or perhaps interacting ghost scalar fields may stabilize the system, and we emphasize that the situation is highly non-trivial. Work along these latter lines is presently underway.

\section*{Acknowledgements}
FSNL acknowledges financial support of the Funda\c{c}\~{a}o para a Ci\^{e}ncia e Tecnologia through the grants CERN/FP/123615/2011 and CERN/FP/123618/2011.

\end{document}